\documentstyle[12pt]{article}

\newcommand{\be}{\begin{equation}}
\newcommand{\ee}{\end{equation}}
\newcommand{\bea}{\begin{eqnarray}}
\newcommand{\eea}{\end{eqnarray}}

\begin{document}

\begin{center}
\begin{large}
{\bf  How {\it Not}  
  to Construct \\}
{\bf an Asymptotically de Sitter Universe \\}
\end{large}  
\end{center}
\vspace*{0.50cm}
\begin{center}
{\sl by\\}
\vspace*{1.00cm}
{\bf A.J.M. Medved\\}
\vspace*{1.00cm}
{\sl
Department of Physics and Theoretical Physics Institute\\
University of Alberta\\
Edmonton, Canada T6G-2J1\\
{[e-mail: amedved@phys.ualberta.ca]}}\\
\end{center}
\bigskip\noindent
\begin{center}
\begin{large}
{\bf
ABSTRACT
}
\end{large}
\end{center}
\vspace*{0.50cm}
\par
\noindent

\par
Observational evidence suggests that  our 
universe is currently evolving  towards  an asymptotically
de Sitter future.  Unfortunately and in spite
of much recent attention, 
various quantum, holographic and cosmological
aspects of  de Sitter space 
remain quite enigmatic.  With such intrigue in mind, this paper 
considers  the ``construction'' of a toy model
that describes 
 an asymptotically de Sitter
universe. More specifically, 
we   add
fluid-like  matter 
to an otherwise purely de Sitter  spacetime, formulate
the relevant  solutions
and then discuss the cosmological and holographic implications.
If the objective is to construct an
asymptotically de Sitter universe that
is free of singularities and has a straightforward
holographic interpretation, then the
results of this analysis are decidedly negative.
Nonetheless, this toy model 
nicely illustrates the pitfalls
that might be encountered in a more realistic
type of construction.

\newpage

\section{Introduction}

Recent astronomical  observations have implied that the
physical universe is currently in a   phase of acceleration \cite{bops}. If 
this acceleration happens to be  eternal, then our universe
must eventually exhibit a cosmological event horizon.
Furthermore, the universe
 will likely face an asymptotically de Sitter
future; that is, a future spacetime that
is dominated by a fixed, positive cosmological
constant. It should be pointed out, however, that 
alternative scenarios may still  account for  this
accelerating  phase. For instance, rather
than a cosmological constant {\it per se},
  a dilatonic scalar field rolling
in a stable potential towards a vanishing minimum; also known as
``quintessence''  \cite{q1,q2}.  Nevertheless, some  recent studies
 have demonstrated \cite{hksx,fkmpx}    that the problematic
features of an asymptotically de Sitter future,
as discussed later in the  section, do indeed persist
in these alternative cosmologies.
\par
 The empirical
evidence of an accelerating universe, as well as general curiosity,
has provoked a recent flurry of  research activity  
 in the realm of  de Sitter (dS) space and, more generally, asymptotically
de Sitter (AsdS) spacetimes. In particular,
holography \cite{tho,sus} in a de Sitter  setting
has sparked significant interest. (See Ref.\cite{med2}
for a list of relevant citations, as well as
Refs.\cite{mci,bmsx,svx,caix,halx,myux,onx,dasx,bckzx,lmmx,nsx,lvl,blah2,
dls,med,blah,klxxx} for particularly recent work.)
However, in spite of all this attention, many aspects of  AsdS spacetimes
remain quite  enigmatic.  As a consequence,  
it has often been conceptually 
 difficult  
to interpret (i) string theory, (ii)  holography  and/or
(iii) physical  observables 
in a de Sitter-cosmological framework.
Next, let us  briefly discuss some of the more perplexing issues;
with the discussion  categorized,  perhaps arbitrarily, 
in terms of the above list of topics.
\\
\par
{\it (i) String Theory}:\footnote{For
a  review  on string theory and M-theory 
that is aimed at the ``lay-physicist'',
see Ref.\cite{polxxx}. Note that, in the current discussion,
our usage of  string theory typically implies M-theory as well.}   
Although a definitive theory of quantum gravity is still lacking,
the consensus  of opinion  suggests that string theory -
and/or  its co-conspirator, M-theory -  will have  a
 significant role to play in the ultimate, fundamental theory.
Unfortunately, string theory, as we currently understand it,
does not appear to be
  compatible with de Sitter space.  To explain this apparent impasse,
let us consider the following.  
String theory is, of course, an inherently
supersymmetric construction; thus implying a vanishing cosmological
constant, as long as  supersymmetry (SUSY) is preserved.
This observation is, in itself,  
 not problematic,  considering that   SUSY
is clearly broken in  our physical universe.  So  the challenge
is really to find a SUSY-breaking string-theoretical vacuum  
that leaves
us with an approximately flat,  positively curved
spacetime (as dictated by observational evidence).
\par 
Alas, any attempt in the  prescribed direction  has   failed to
provide an acceptable description of  our physical reality. 
Firstly, in a context of perturbative or weakly coupled string theory,
SUSY-violating vacua  tend to have tachyonic
instabilities. These, in turn,  either
give rise  to a large, negative vacuum energy or drive the 
theory into an unphysical regime of decoupled gravity \cite{banks}.
Secondly, in a  context of 
 non-perturbative string theory  or matrix
theory \cite{matrix},  any relevant  vacuum collapses into a 
singular spacetime  when SUSY is
broken.  Technically speaking, this collapse can be viewed 
as a consequence of a pertinent ``no-go'' theorem 
\cite{nunmal}.\footnote{There
have been many interesting ways of circumventing this type of
 no-go theorem; 
however, at the expense of problematic features such as
wrong-sign kinetic terms and non-compact ``compactification'' manifolds.
See Ref.\cite{dds} for related discussion and references.}
In a more fundamental sense, this failure  can probably
be attributed to de Sitter space having a finite number
of degrees of freedom (see below);  in complete contrast
to matrix theory's  implication of  an infinitely large Hilbert space.
\par
{\it  (ii) Holography}:  The holographic principle, in its most
basic sense,  
  places an upper  limit on the amount of information
(i.e., entropy)
that can be stored in a given region of spacetime \cite{tho,sus}. 
The generality of this statement can  readily  be seen
by way of  a covariant entropy bound that  was 
first proposed by Bousso \cite{bound}.  In particular, this bound
directly relates 
the area of a given  spatial surface 
to the maximal  entropy passing through  appropriately 
generated light sheets. 
Applying this formalism, Bousso went on to demonstrate
that the entropy of pure de Sitter space  will serve
as an upper bound on the ``accessible'' entropy\footnote{Accessible 
versus inaccessible
 entropy essentially
makes the distinction between  those degrees of freedom that
can or cannot  influence (and be influenced by)
a given experiment.} in any
AsdS spacetime \cite{bou}. 
\par
As previously indicated, this finite entropic bound 
implies an incompatibility between
 string (or matrix)  theory and  de Sitter space.
The same  basic difficulty arises in the (conjectured) holographic
duality between an $n$+1-dimensional de Sitter bulk  
and an  $n$-dimensional  
conformal field theory (CFT) \cite{str}.\footnote{If valid, such a 
dS/CFT duality
follows, by way of analogy,  from  the highly
successful anti-de Sitter/CFT  correspondence \cite{mal,gub,wit}.}
That is to say,  the  holographic dual of an AsdS spacetime - a Euclidean
 CFT that ``lives'' on  a spacelike asymptotic boundary -  is
described by an infinite-dimensional  Hilbert space and 
(therefore) has access to an infinite number of  degrees of freedom. 
Equally  distressing, the  entropic upper bound seems to
imply (at least indirectly via a closely related mass bound \cite{bdm})
that the dual boundary theory is a non-unitary one \cite{cai,hal,med2}.
\par
While considering the paradoxical implications of the entropy bound, one
should keep  in mind that 
 the holographic principle 
does {\it not}, on its own accord, 
prescribe the existence of a dually related boundary theory 
\cite{bouxx}.\footnote{In this regard,
holography can be viewed as a necessary but {\it insufficient} prerequisite
for establishing the anti-de Sitter/CFT correspondence.} 
 But,  on the other hand, the evidence
in favor of  a de Sitter-based duality does indeed appear  to be mounting
(see the previously cited literature).   However, for
a very  recent argument  in opposition to  a viable 
dS/CFT correspondence, see Ref.\cite{dls}. 
\par
{\it (iii) Physical Observables}:  
The feasibility of any proposed cosmology depends
on the existence of  well-defined 
observables that  intelligent  ``passengers''   
can make sense of.  For sake of  argument, let us consider
a (very) hypothetical asymptotically flat universe. 
In this case, the physical
 observables can be defined (and only defined) in terms of
the S-matrix elements of asymptotically free particle states.
An asymptotically anti-de Sitter universe provides
a similar picture, except that the S-matrix  is replaced by
the boundary correlators of bulk fields \cite{mal,gub,wit}.
What do both of these hypothetical cosmologies have in common?
Precisely that there is an infinitely large asymptotic boundary
(at spatial infinity)
where the bulk degrees of freedom  can separate into free particles.
\par
With the above discussion in mind,
we can now see  that  observables are, at best, poorly defined
in  an   AsdS cosmology (and, in fact, any
cosmology that admits a causal event horizon \cite{q1,q2}).  
 First of all,  there is 
no spatial  infinity to speak of.  There is, at least,
a future infinity; 
however,   for any given observer, this is nothing but a singular
point.  That is,  an observer's causal future shrinks to nothingness
in the asymptotic limit.
To put it another way, any two events that are sufficiently
close   to   temporal infinity   will  have no common future and
(therefore) immeasurable correlations.   Consequently,
any  related S-matrix-like construction  should   have no physical
 meaning.
\par
With regard to the above conundrum, Witten  has suggested \cite{witxx}
that one can still construct a sort of ``S-vector''; the elements
of which measure the probability of a  given  initial state (at
past infinity)
 to evolve into a given  final state (at future infinity).
This would allow the calculation of quantities that he refers
to as ``meta-observables''.  However,   it is
not entirely clear as to  what type of observer could make
sense of  meta-observables, insofar  as it would require
a global perspective that is outside the realm of  ``common''
passengers.  In another interesting, related discussion,
Banks and Fischler \cite{bfxx} have proposed an S-like matrix that
interpolates between  a unique initial state (presumably, a
``big bang'') and the states on an observer's  cosmological horizon.
\\
\par
With  the study of AsdS cosmologies
 having sufficiently been motivated, we now  focus our attention  on
the upcoming analysis.  However,   before discussing
the actual content of the current program,  let us
first consider a recent, topical paper  by Leblond {\it et al}
\cite{lmmx}. This informative work covered many  aspects of de Sitter space;
one of which was the ``construction'' of an  AsdS spacetime
containing non-trivial bulk fields.  In particular, this ``matter'' 
was formulated  in terms of a dilatonic scalar, with the  gravity-dilaton 
coupling  being described by a  
rather exotic potential.\footnote{Actually,
 the
authors of Ref.\cite{lmmx} considered a pair of related potentials:
 a discontinuous
(``step'') potential and its smooth-functional ``analogue''. The latter form
is more realistic but
necessitated a numerical analysis.} One might wonder
if a similar construction can be achieved  in terms
of more ``conventional'' matter. This, in a nutshell,
is the premise that underlies the current treatment.
\par
To keep things simple (and thus calculable),
we  limit considerations to a toy  model  for
a ``perturbed'' de Sitter cosmology. 
More specifically,  fluid-like matter  is added
to an otherwise purely de Sitter spacetime (of arbitrary dimensionality). 
We also constrain the matter to have a fixed
equation of state and to satisfy the  ``causal     
energy condition'' \cite{wald}.
We do, however, consider matter with    both
a  positive and  negative energy density.   
The latter, somewhat unorthodox choice will be rationalized at an appropriate
juncture.
\par
Let us point out that the methodology employed here
closely follows that of McInnes \cite{mci}. In fact,
this author considered the specific case of negative-energy
matter (in a 4-dimensional AsdS spacetime);  
the purpose  being to investigate
the feasibility of  so-called ``phantom'' cosmologies \cite{phantom}.    
Hence, with regard to the negative-energy case, there is some overlap 
 between the two studies.  Nevertheless,  we feel
that our perspective
and  interpretations are  somewhat different.  
\par
The remainder of the paper is organized as follows.
In Section 2, we introduce the  pertinent 
formalism  and establish  the appropriate boundary conditions;
enabling a precise description of     the cosmological model
of interest
(see  above).  Early on, a distinction is made between
a pair of very different  cosmological scenarios: an AsdS spacetime
 perturbed with  
($A$) positive-energy matter  and ($B$) negative-energy matter. 
In Section 3, we explicitly formulate the spacetime 
metric for both Case $A$ and $B$.  After which,  the results
are interpreted and  various  implications are discussed.
In Section 4, we reconsider the prior outcomes 
from a holographic point of view. In particular,
we employ  a standard prescription 
\cite{str2,bdm}
to calculate generalized $C$-functions and 
then study the associated renormalization group flows 
(for instance, \cite{ag,fgpw,sak,dvv,str2,bdm,lmmx}).
Finally, Section 5 contains a summary and concluding remarks.
\par
To cut to the chase, we find   both  cosmological models of interest
to be problematic.  The positive-energy case  contains
a true (curvature)  singularity, whereas  the negative-energy case
 may have irreconcilable 
difficulties from a holographic  perspective. 
(This latter ``breakdown'' 
was first pointed out in
  Ref.\cite{mci}; however,  unlike the prior work, we argue that
this result   may be open to interpretation.)
 Let us, therefore, emphasize that  our current intention
 is {\it not} to construct a viable AsdS
cosmology.  Rather, we hope  to  illustrate, by way of a toy model, 
the pitfalls
that  may be encountered  when and if    a  more realistic
 construction is  attempted.

\section{Preliminaries}

We  begin the formal analysis by considering an $n$+1-dimensional
Friedmann-Robertson-Walker (FRW) spacetime with a scale
factor of $a(t)$ (where $t$ is the cosmological time).
The corresponding FRW metric can be expressed
by way of the following line element:
\be
ds^2=-dt^2+a^2(t)d\Sigma^2_{n,k}.
\label{1}
\ee
Here, $d\Sigma^2_{n,k}$  describes the $n$-dimensional (Euclidean)  metric of
a constant-time hypersurface, such that
 $k$  parametrizes  the curvature.
For spherical, flat or hyperbolic spatial sections,
$k$ respectively takes on a value of $+1$, $0$ or $-1$.
\par
Let us assume, for sake of  simplicity, that
  the spacetime is filled with  a perfect fluid of energy density
$\rho$ and pressure $p$.\footnote{A perfect fluid insinuates
that the stress-energy tensor is diagonalized according to
$T^{\mu}_{\nu}=diag\left[-\rho,p,p,...,p\right]$.} 
For matter such as this,
 it is known that
Einstein's field equations
($G_{\mu\nu}=R_{\mu\nu}-{1\over 2}Rg_{\mu\nu}=T_{\mu\nu}$)\footnote{Note
that the cosmological constant, if non-vanishing, is assumed to
be absorbed into the stress-energy tensor.}
 take on a Friedmann-like form  (for instance, \cite{towly}).
More specifically, one finds that the Einstein
tensor equation  reduces to  the following
 pair of    expressions: 
\be
\left( {{\dot a}\over a}\right)^2+{k\over a^2}={16\pi G\over n(n-1)} 
\rho,
\label{2}
\ee
\be
{{\ddot a}\over a}=-{8\pi G\over n (n-1)}\left[(n-2)\rho+n p\right],
\label{3}
\ee
where $G$ is the $n$+1-dimensional Newton gravitational constant and
a dot indicates differentiation with respect to $t$.
\par
After  some manipulation,  the above field equations can
be utilized to derive  the following energy-conservation 
relation:
\bea
{\dot \rho} &=& -n{{\dot a}\over a}\left[\rho +p\right]
\nonumber \\ 
&=&  -n{{\dot a}\over a}\left[1 + \omega\right] \rho.
\label{4}
\eea
In the lower line, we have introduced the usual
 equation-of-state parameter,
$\omega$, such that $p=\omega\rho$. Some  well-known (fixed) 
choices for $\omega$
include  a universe that is purely filled with  radiative matter
 ($\omega=1/n$), 
dust-like  matter ($\omega=0$), 
a positive  cosmological constant ($\omega=-1$ with
$\rho > 0$),
and a negative cosmological constant ($\omega=-1$ with $\rho < 0$). 
Although, generally speaking, one could always regard
$\omega$ as a function of space and time.  Note that 
$\left|\omega\right|\leq 1$ is  often  enforced, which translates to the 
causal energy condition 
\cite{wald}. However, for sake of generality, this will not
necessarily be the case  in the analysis to follow.\footnote{That is 
to say,
the overall stress tensor (including cosmological constant) will not 
necessarily be subjected to this causality condition; however,
the  matter  will be.}
\par
Since our current focus is on  ``perturbed''  de Sitter spacetimes,
it is instructive to first consider the above  picture for
pure de Sitter space. In this case of positive cosmological
constant (i.e., $\Lambda >0$) and no other matter, 
$\omega_{dS}=-1$ and Eq.(\ref{4})
reminds  us that the energy density is a positive constant as well.
For future reference,  we present the de Sitter  line element
in both  planar coordinates ($k=0$) and global coordinates
($k=1$)   \cite{ssv}:
\be
\left. ds^2_{dS}\right|_{k=0}= -dt^2+e^{2t/L}\sum_{i=1}^{n}dx^2_{i}, 
\label{111}
\ee
\be
\left. ds^2_{dS}\right|_{k=1}= -dt^2+ \cosh^{2}\left({t\over L}\right)
L^2\left[d\chi^2+\sin^2(\chi)d\Sigma^2_{n-1,1}\right],
\label{222}
\ee 
where $L$ is
the  curvature radius defined  by $\Lambda=n(n-1)/2L^2$. 
Note that the planar coordinate system  only describes half of the 
de Sitter
manifold, but one can obtain the other half
by  substituting   a minus sign into the exponential.
\par
Sometimes, it is more instructive  when de Sitter space is 
expressed in conformal coordinates.
We do so here for the global ($k=1$) case:
\be
\left. ds^2_{dS}\right|_{k=1}= {L^2\over \sin^2(\eta)}
\left[-d\eta ^2 +  d\chi^2+ \sin^2(\chi)d\Sigma^2_{n-1,1}\right],
\label{333}
\ee 
where $d\eta=dt/L\cosh(t/L)$. With a suitable choice of
integration constant, this translates into
  $\eta(t)=2\tan^{-1}\left[e^{t/L}\right]$.   
Take note of the finite extent of the conformal (i.e., Penrose)
diagram. It has a ``height'' of $\Delta\eta =\eta(+\infty)-\eta(-\infty)
=\pi$ and a ``width'' of $\Delta\chi=\pi-0=\pi$.
Also of interest, any observer must suffer  the indignities
of both a past and future cosmological  horizon.
For example, an observer  at $\chi=0$
has a future horizon at $\chi=\pi-\eta$ and a past horizon at 
$\chi=\eta$. For further background on de Sitter space and
 plenty of fancy diagrams,  one
can consult Refs.\cite{ssv,lmmx,bouxx}.
\par
An interesting feature of de Sitter space is that, regardless of
the spatial slicing  or choice of $k$, one finds that 
$a(t)\sim e^{t/L}$ as $|t|\rightarrow\infty$ \cite{lmmx}.
We can use this  observation, along with Eq.(\ref{2}),
to identify the (constant)
 energy density and pressure of  a  de Sitter cosmology:
\be
\rho_{dS} = - p_{dS} = {n(n-1)\over 16\pi G L^2}.  
\label{4.5}
\ee
\par
Let us now concentrate on the scenario of interest; namely,
incorporating matter (to be denoted by $\phi$) into an otherwise
purely   de Sitter spacetime. 
The
only stipulations on  this supplementary  matter will
be that it 
is of a perfect-fluid
form (so that Eqs.(\ref{2}-\ref{4}) remain in effect),
it has a fixed   equation of state  
(i.e., $\omega_{\phi}=p_{\phi}/\rho_{\phi}$ is a constant), 
  and it obeys the 
causal energy bound (i.e.,   
 $\left|\omega_{\phi}\right|\leq 1$).\footnote{This causality
condition ensures that  energy travels slower than light
when such matter serves as the medium \cite{wald}.}
The total energy and  pressure are thus defined as follows:
\be
\rho=\rho_{dS} + \rho_{\phi},
\label{4.6}
\ee
\be
p= p_{dS} + p_{\phi} =-\rho_{dS}+\omega_{\phi}\rho_{\phi}.
\label{4.7}
\ee
It should be kept in mind that the de Sitter contributions
cancel off in the  summation of  $\rho$ and $p$. 
\par
As implied by the above discussion, we will  permit
$\rho_{\phi}$ to be  positive or negative; although the latter
situation  is
in conflict with the (usually enforced) ``weak energy condition'' \cite{wald}.
If some readers are bothered by such an exotic choice of matter,
it may help to think of negative  $\rho_{\phi}$ as describing the removal
of energy (from pure de Sitter space)  as opposed to the 
addition of negative energy. Indeed, a  Schwarzschild-de Sitter
black hole spacetime (for example) 
is  really just  a negative  energy excitation
of pure de Sitter  space under static, spherically symmetric conditions 
\cite{bdm}.
\par
The  working hypothesis will be a universe that  is asymptotically  de Sitter
(AsdS) in 
the distant past ($t = -\infty$) and, barring curvature singularities or
divine intervention, evolves into an AsdS spacetime in the  far
future ($t=+\infty$).  What will  become quite  evident, as the analysis
proceeds, is   two  distinctive   solution sets
depending on the positivity/negativity of 
$\rho_{\phi}$.\footnote{By way of Eq.(\ref{4}), the causality bound,
the AsdS initial condition and  time-reversal
symmetry, it can be shown  that
 $\rho_{\phi}$  does  not change its sign  during the history of 
the universe.}  
To help clarify matters, we  thus establish the following
classifications: \\
i)  Case $A$ for   $\rho_{\phi} > 0$ (hence, $\rho+p >0$); \\
ii) Case $B$ for   $\rho_{\phi} < 0$ (hence, $\rho+p <0$).  \\
We take this opportune moment to (again) point out  that
a prior, related work by McInnes \cite{mci}
 studied  the cosmology  corresponding
to  Case $B$  with  $n=3$ and $\omega_{\phi} <0$.
\par
Given our working hypothesis of an AsdS spacetime, it follows
that we can write:
\be
\left.\rho\right|_{A,B} = \rho_{dS} \pm S(a)
\label{6}
\ee
where $S(a)$ is some well-behaved, positive
 function that satisfies
$S(a)\rightarrow 0$ as $a^2$ (or $|t|$) 
$\rightarrow\infty$. Note that, where applicable,  the left/right
subscript  label  corresponds to the  upper/lower sign.
Also note that $S(a)$ is not {\it a priori} the same function  
in the two different cases. 
\par
At this point in the analysis, it proves   convenient to 
  define the following parameter:
\be
\gamma \equiv n(1+\omega_{\phi}).
\label{7}
\ee
Note that, strictly speaking, $0\leq \gamma \leq 2n$.  
Although, if the resultant spacetime 
is to be truly classified as a ``perturbation'' of de Sitter
space,  $\gamma$ should really be  limited
to values much  smaller than one. As it turns out,
this distinction is never really an issue. 
(Contrary to this statement is, however, the last topic of Section 3.)
\par
Given the above,
Eq.(\ref{4.7}) for
the total pressure can now be re-expressed as follows:
\be
p=-{\gamma\over n}\rho_{dS}-\left[{n-\gamma\over n}\right]\rho.
\label{10}
\ee
Further incorporating Eq.(\ref{6}), we have:
\be
\left. p\right|_{A,B}=- \rho_{dS}\mp\left[{n-\gamma\over n}\right]S(a)
\label{11}
\ee
or, more to the point:
\be
\left.\rho+ p \right|_{A,B}= \pm {\gamma\over n}S(a).
\label{11.5}
\ee
\par
For  Case $A$ and $B$ alike,
   the above outcomes  can be substituted into
the energy-conservation equation (\ref{4}) to yield:
\be
{\dot S}=-\gamma{{\dot a}\over a} S(a).
\label{12}
\ee
Therefore:
\be
{dS\over da}=-{\gamma\over a}S(a).
\label{13}
\ee
This differential equation  can be
 trivially solved as follows:  
\be
S(a)=\alpha a^{-\gamma},
\label{14}
\ee
where $\alpha$ is a positive, dimensional constant. 
Note that the asymptotic behavior
of $S(a)$ is precisely  as anticipated.

\section{Perturbed de Sitter Space}

Our next objective will be  to utilize the prior formalism
(in particular, the most recent outcome and the field equations)
 in obtaining
 an explicit   solution for the scale factor, $a(t)$.
To substantially simplify this task, we will follow
Ref.\cite{mci} and assume that the spatial sections
are exactly flat (i.e., $k=0$) and compact.
Although this choice is  motivated, in large part, by convenience,
let us point out that (approximate) flatness  follows  from 
 observational evidence
and  inflationary implications  
 \cite{carr}; whereas compactness
 can be argued for by way of the following
 holographic consideration.
\par
In regard to the argument
for compactness, first recall (from  Section 1) that
there is a holographic upper limit on the number of 
{\it accessible} degrees of freedom  in any AsdS spacetime 
\cite{bou}.  Now
consider that an arbitrary perturbation of  de Sitter space will
typically allow an entire Cauchy surface to be  visible   
 at some finite value of  time  \cite{gao,lmmx}.
It is clear that, at least naively, these two statements
are contradictory if the  volume 
of  the relevant Cauchy surface  can  increase without bound.
Thus, compactness of the spatial slices  naturally follows.
\par
In any event, the  intricacies of the spatial slicing
will have  no repercussions on the de Sitter picture at large
values of $|t|$, where   such information is ``conveniently'' concealed
behind a cosmological horizon.\footnote{Any future life-form
trying to cope with  the  bleak final stages of
an AsdS cosmology  \cite{witxx}
would likely argue that this information loss
is anything but  convenient.}  
\par
For sake of definiteness (and roughly following \cite{mci}), let us 
regard the spatial sections as cubic tori that are
 parametrized by $n$ angular coordinates, $\theta_i$.
That is, the FRW metric (\ref{1})   now takes on the
following form:
\be
ds^2=-dt^2 + a^2(t) {\cal A}^2 \sum^{n}_{i=1}d\theta_i^2 ,
\label{17}
\ee
where ${\cal A}$ is some  arbitrary length parameter.
\par
We now return to the issue at at hand - the evaluation
of  the scale factor -  and  reconsider the first-order  field
equation  (\ref{2}). Also incorporating Eqs.(\ref{4.5},\ref{6},\ref{14})
and  setting  $k=0$, we have:
\be
\left({{\dot a}\over a}\right)_{A,B}^2={1\over L^2}\pm {16\pi G \over
n(n-1)}\alpha a^{-\gamma}.
\label{18}
\ee
\par
The above equation can readily  be solved, provided that  
  $\alpha$ has been identified with $n(n-1)/16 \pi G L^2$. 
Accepting this quite reasonable  identification,
we find (for Case $A$ and  $B$, respectively):
\be
\left. a(t) \right|_{A}=
\sinh^{\left(2\over \gamma\right)}\left({\gamma (t-t_{o})\over 2L}\right),
\label{19}
\ee
\be
\left. a(t) \right|_{B}=
\cosh^{\left(2\over \gamma\right)}\left({\gamma (t-t_{o})\over 2L}\right).
\label{20}
\ee
Note that $t_o$ is a constant of integration, which, without loss
of generality, 
will subsequently be set to vanish. 
\par
We can now  display  the respective solutions with their
time dependence explicitly indicated: 
\be
ds^2_{A}=-dt^2 +{\cal A}^2 \sinh^{\left({4\over \gamma}\right)}
\left({\gamma t\over 2L}\right)
\sum^{n}_{i=1}d\theta_i^2 ,
\label{21}
\ee
\be
ds^2_{B}=-dt^2 +{\cal A}^2 \cosh^{\left({4\over \gamma}\right)}
\left({\gamma t\over 2L}\right)
\sum^{n}_{i=1}d\theta_i^2 .
\label{22}
\ee
\par
As a check on consistency, let us examine the asymptotic behavior
of these solutions.
 Setting  $x_i=2^{-2/\gamma}{\cal A}\theta_{i}$
and  taking  $|t|>>1$, we obtain the following  expression 
in both cases: 
\be
ds^2 \approx -dt^2+e^{2|t|/L} \sum^{n}_{i=1}d x_i^2 .
\label{23}
\ee
Notably,  this asymptotic form agrees with Eq.(\ref{111}), which
describes 
  a purely  de Sitter spacetime
with flat spatial slicing.  Moreover,
this  form  is, as previously discussed, the
unique FRW description of any de Sitter space 
in the limit of  
large $|t|$  \cite{lmmx}.  Thus,  our toy model
displays
 the anticipated AsdS behavior, regardless of the sign of
the energy density.
\par
Next, we comment on some of  the implications 
of these solutions, starting with Case $A$.
It is quite  clear that the corresponding  scale factor (\ref{19})  has a zero
on the constant-time surface defined by  $t=0$. What is not yet so evident 
is the physical significance of this zero. That is to say,  does it represent:
(i) a singularity  in the spacetime beyond which time can not flow
or (ii)  a Cauchy null surface beyond which the solution can (and should)
be analytically continued? The simplest way to  resolve this
dilemma is to consider the scalar curvature on the surface in question.  
Calculating the
prescribed quantity, we find:
\be
\left. R \right|_{A} 
={n(n+1)\over L^2}+{n\over L^2}\left[n+1-\gamma\right]
{1\over \sinh^2\left({\gamma t\over 2 L}\right)}
\label{25.2}
\ee
and, for completeness:
\be
\left. R \right|_{B}
={n(n+1)\over L^2}-{n\over L^2}\left[n+1-\gamma\right]
{1\over \cosh^2\left({\gamma t\over 2 L}\right)}.
\label{25.3}
\ee
\par
We see that, for Case $A$, the scalar curvature (\ref{25.2})
``blows up'' at $t=0$.
Hence, the corresponding  zero in the metric
does indeed represent a true singularity in the fabric
of spacetime.
 Physically,
this outcome can be perceived as a  ``big crunch''  at which  the  universe
collapses into  a singular point.\footnote{Alternatively, by
way of time-reversal symmetry (as  implied in this toy model),
one can just as easily regard this singularity as a ``big bang''.
The choice of bang versus crunch  depends on whether  the singularity 
precedes 
or proceeds  time evolution. Until one invokes the second
law of thermodynamics  to  point the way, 
it is really just a matter of taste.}
Moreover, we can, at least conjecturally,  view
  this outcome as a  manifestation of
the AsdS entropy bound \cite{bou}. To re-clarify,
it has been demonstrated that the entropy of pure de Sitter
space ($\sim L^n/G$) should  serve as an 
upper bound on the number of 
 degrees of freedom that can  experimentally be probed in an 
AsdS spacetime.\footnote{In qualifying the degrees of freedom, 
we are making the
distinction between those that are accessible versus inaccessible.  
  Accessible degrees of freedom  can  both influence and
be influenced by a given experiment. That is, the experiment in question
determines a ``causal diamond'' or accessible domain
of the spacetime \cite{bou}.}    
With the addition of positive-energy matter 
to an otherwise purely de Sitter spacetime,  it follows,
at least naively, that this
entropic upper bound has somewhere been violated.  Consequently,  any
chance of a final  AsdS phase  has  effectively been thwarted. 
\par 
With regard to the singular nature of Case $A$, 
let us further point out the following. Since
this cosmological scenario  describes the addition of (positive-energy)
 matter, rather than
entropy  {\it per se}, 
 it is (perhaps)  most appropriate
to interpret the singularity as  a violation of a proposed
 AsdS mass bound \cite{bdm}.\footnote{On the other hand,
the entropy \cite{bou} and mass \cite{bdm} bounds are closely related and
may actually  represent different semantics for the same
physical principle.}  
  To elaborate, it has recently been argued that
   the mass of  pure de Sitter
space should serve as an upper bound on the total mass
of any AsdS spacetime. Moreover,  the authors of Ref.\cite{bdm} 
have stressed that a violation of this bound
should induce a cosmological singularity.  Notably,
this last point  is clearly  supported by our findings. 
\par  
Let us now consider Case $B$, for which  the plot changes
dramatically.  For this scenario, the  scale factor  (\ref{20}) and 
scalar curvature (\ref{25.3}) are both clearly
regular throughout the spacetime manifold. Therefore,
a perturbed de Sitter universe that remains AsdS, in both
the distant past and the far future, has truly been realized. 
This outcome is not particularly surprising for two
reasons. First of all, technically speaking,
the AsdS entropy bound \cite{bou} only has validity when the
null  energy condition \cite{wald}
has been satisfied.
(To translate, this condition, along with causality, forbids negative-energy
matter.) Secondly, on a more intuitive level,
this negative-energy case can  effectively
be viewed as a removal of positive-energy matter from an
otherwise purely de Sitter  spacetime.  
That is, Case $B$  translates into, if anything, a reduction
in the number of accessible degrees of freedom.  Thus, from
the viewpoint of either relevant bound (mass \cite{bdm} or
entropy \cite{bou}),
there is  nothing to obstruct this cosmology
from persevering into the far future.
\par
As a topical aside, we point out that Case $B$ provides 
a concrete realization of the so-called 
``tall'' and ``wide'' universes that have recently been
advocated  by Leblond {\it et al} \cite{lmmx}.
To make a tall-story short, Gao and Wald \cite{gao} have 
demonstrated that a (reasonably) generic perturbation of
de Sitter  space leads to an increase  in the 
 height of its  conformal diagram (see the discussion
following Eq.(\ref{333})).  
This  suggests that,
given a perturbed but  singularity-free de Sitter spacetime, 
 a complete 
Cauchy surface will become entirely visible
to an appropriate  observer at a finite value of time. One can further envision
an AsdS universe that is not only tall (in the above sense)  but
also arbitrarily wide; that is, the spatial slices and (therefore)
visible Cauchy surface  could  well have  an arbitrarily
large volume. 
\par
How does the above discussion apply to the spacetime described by
Case $B$?  As it so happens,  one can make the  full extent
of conformal time  arbitrarily
large by taking   $\gamma$ to be arbitrarily
small \cite{mci}. (Recall that 
  $0\leq \gamma \leq 2n$ is the only formal stipulation and we
ethically  prefer a small value of $\gamma$ in any case.)
To substantiate this claim,
let us first  rewrite Eq.(\ref{22}) in terms of conformal
coordinates:
\be
ds^2_{B}= L^2 \cosh^{\left({4\over \gamma}\right)}
\left({\gamma t\over 2L}\right)
\left[-d\eta^2 +\left({{\cal A}\over L}\right)^2
\sum^{n}_{i=1}d\theta_i^2\right].
\label{99}
\ee
 The full  extent of conformal time is then
given by:
\be
\Delta \eta= {1\over L}\int^{+\infty}_{-\infty}
{dt\over \cosh^{\left({2\over\gamma}\right)}\left({\gamma t\over 2L}
\right)} ={2\over \gamma}\int^{\pi}_{0}\sin^{\left({2\over\gamma}-1\right)}
(\xi)d\xi.
\label{25.1}
\ee
\par
Even without an explicit evaluation of this integral, 
it is sufficiently evident that $\Delta\eta$
can be made arbitrarily large  in the manner  prescribed above.
Furthermore, the spatial sections, although described as compact,
can be made to have an arbitrarily large (but still finite) volume
with a suitably large choice for  ${\cal A}$.
Thus,  we can arrange  the  Case-$B$ universe  to be  both tall and   
wide enough to fulfill anyone's  expectations, no matter how
greedy. What makes this outcome somewhat of a  puzzle is 
that the analysis  of Gao and Wald \cite{gao}, which
conceptually presupposes the tall/wide construction \cite{lmmx},
 only applies 
to  perturbations  satisfying  the null  energy condition.
Case $B$ is, by hypothesis, clearly in violation of
this condition, and we presently have no explanation for
this phenomenon.

\section{RG Flows}

As it is well known,  there are many indicators
that AsdS spacetimes can be  holographically described 
 by an appropriate  conformal field theory (CFT).
In particular, this dually related CFT is
  a Euclidean one  that
lives on a spacelike asymptotic  boundary of
the  bulk spacetime \cite{str}.
(Consult Section 1 for further references.)  
If valid, such a dS/CFT duality follows, by way of analogy,
from the celebrated anti-de Sitter (AdS)/CFT 
correspondence \cite{mal,gub,wit}.
It remains unclear, however, as to what extent   
anti-de Sitter  holographic features can be generically extrapolated
into the  de Sitter-based  duality.\footnote{For
a recent critique on the dS/CFT correspondence, see  Ref.\cite{dls}.}
Nonetheless, many such features  do indeed persevere
into a de Sitter realm; including  the intriguing phenomena
of  renormalization group (RG) flows.  
\par
Before proceeding to the analysis, let us discuss the premise behind  
holographic  RG flows.\footnote{See, for
instance, Refs.\cite{ag,fgpw,sak,dvv,str2,bdm,lmmx} for further, 
pertinent discussion.}
Firstly, from an AdS/CFT perspective,
it has been established  that 
the monotonic evolution of a relevant    bulk parameter induces 
 a  ``flow''  in the
renormalization scale of the boundary theory  and {\it vice versa}.
(Note that this renormalization scale determines
the ultraviolet cutoff or lattice spacing  of the dual  CFT.)
This picture  follows, in large part,  from
the  ultraviolet/infrared (UV/IR) correspondence  \cite{sw,pp}.
That is,  large distances (IR) in the anti-de Sitter bulk 
correspond to  small distances (UV) on the conformal boundary 
and {\it vice versa}.
\par
It should be kept in mind that
a   holographic RG flow depends on the existence of
a generalized $C$-function.  This follows, by way of analogy,
 with  $C$-functions 
 in a two-dimensional CFT context \cite{zam}.   
For  AdS/CFT  holographies  in particular,
 the  generalized  
 $C$-function should, if properly identified,  exhibit various
monotonicity properties that are reflective
of the underlying UV/IR duality \cite{sw,pp}. 
\par
Next,  let us consider RG flows  in  a
 dS/CFT  holographic context.
As observed by Strominger \cite{str2} (also see \cite{bdm}),
time evolution  in a purely de Sitter bulk will
generate conformal-symmetry  transformations  on its asymptotic boundaries
(spacelike past infinity and future infinity).
Particularly significant to prospective RG flows, 
  this  conformal symmetry will be  jeopardized 
when the bulk spacetime is ``only'' an  AsdS one.
In this regard,  
it was argued \cite{str2} that AsdS  time evolution will naturally
induce a RG flow between a pair of conformal fixed points.
These fixed points occur in the asymptotic past and
future, where the symmetries of pure de Sitter space ultimately 
emerge. Moreover,  Strominger proposed an  associated $C$-function
of the form \cite{str2}:\footnote{This formulation is identical to
the conventional AdS/CFT $C$-function \cite{ag,fgpw,sak}; except 
that the anti-de Sitter scale factor is a function of radial distance
 rather than time.}
\be
C_{dim=n+1}\sim  \left|\left({{\dot a(t)}\over a(t)}\right)^{-(n-1)}\right|.
\label{88}
\ee
\par
For an AsdS spacetime with flat spatial slicing (and presuming the 
weak energy condition \cite{wald}),
 it has been  shown  that this definition of   $C$  displays appropriate
monotonicity properties \cite{str2,bdm}.\footnote{Recently,
Leblond {\it et al}  have proposed a more generalized
version  of this $C$-function that can be applied to any choice
of spatial slicing \cite{lmmx}. For  our  present considerations,
however, Eq.(\ref{88}) is sufficient.} In particular,
 as time evolves
forward,
$C$  flows to the UV (i.e., increases) 
in an expanding universe and flows to the IR  
 in a contracting
universe. Note that such monotonic behavior is often 
represented  as being  a ``$C$-theorem''. 
(Further note that, for a universe with  both a contracting 
and expanding phase,  one must abandon  the conventional
wisdom of $C$  flowing   in a strict monotonic fashion.  
For further discussion on this caveat, see Ref.\cite{lmmx}.)
\par 
Let us now incorporate the above concepts into
the framework of the current study. We begin here
by considering  positive-energy perturbations of pure
de Sitter space; that is,  Case $A$. Recalling Eq.(\ref{19}) for the 
relevant scale factor, we obtain the following $C$-function
(\ref{88}) (up to irrelevant constant factors):
\be
C_{A}\sim\left|\tanh^{(n-1)} \left({\gamma t\over 2 L}\right)\right|.
\label{30}
\ee
\par
To ascertain the behavior of  $C_A$  as time evolves,
let us consider the following derivative (again
neglecting constant factors): 
\be
{1\over C_{A}}{\partial C_{A}\over \partial t}\sim {t\over |t|}{1\over 
\cosh^2\left({\gamma t\over 2 L}\right)}.
\label{31}
\ee
Upon inspection of this result,
one  can   see  that $C_{A}$ behaves precisley  as anticipated.
To be more specific,  if $-\infty <t<0$ (i.e., a contracting universe),
 ${\dot C}_{A}<0$  and $C_{A}$  flows monotonically to the IR.
Meanwhile,
 if $0<t<+\infty$ (i.e., an expanding universe), ${\dot C}_{A}>0$
and  $C_A$ flows monotonically to the UV.
As a point of interest,
let us  recall that, because  of a curvature singularity, 
Case $A$ can describe
a universe that is   contracting  or  expanding  but 
incapable of  doing  both. Hence,  for any specific model, $C_{A}$  
experiences
a strictly monotonic flow for all relevant  time.
\par
Let us now reconsider Case $B$, describing  
 negative-energy perturbations of de Sitter space.\footnote{Note
that a similar analysis has been carried out by
McInnes in Ref.\cite{mci}.}   In this case,
the appropriate  scale factor is given by Eq.(\ref{20}), which leads 
 to the following results:
\be
C_{B}\sim\left|\coth^{(n-1)} \left({\gamma t\over 2 L}\right)\right|,
\label{33}
\ee
\be
{1\over C_{B}}{\partial C_{B}\over \partial t}\sim -{t\over |t|}{1\over 
\sinh^2\left({\gamma t\over 2 L}\right)}.
\label{34}
\ee
\par
Here, the story  deviates radically  from
what we have found  above. 
For instance, during the contracting phase ($t<0$),
${\dot C}_{B}>0$  and $C_{B}$  flows monotonically to the UV.
Meanwhile,  during the expanding phase ($t>0$),
 ${\dot C}_{B}<0$
and  $C_B$ flows monotonically to the IR.
This behavior  is diametrically opposed to the expectations 
of the conventional  bulk time/RG flow duality. The question  being:
can we provide a physical picture that somehow resolves
this conundrum?
\par
Firstly, let us take the holographic duality literally
and presume that the forward evolution of time
constitutes a
  flow towards the UV/IR
when  the universe is expanding/contracting.
With this strict interpretation,   the  only feasible circumvention
 would be to assume that the universe remains  permanently fixed at $t=0$;
that is,  ``time stands still''.
 On behalf of this viewpoint, we 
point out that the time derivative of the scale factor
vanishes at $t=0$; meaning that this surface represents a
fixed point (infinite UV)  in the holographic dual. 
With regard to this line of reasoning,
it is interesting to recall  our prior analogy between
the Case-$B$ spacetime and a Schwarzschild-de Sitter  black hole. 
To reiterate,  both of these solutions   can be identified
as AsdS spacetimes having  negative energy
relative  to pure de Sitter space. If we extrapolate this identification
to the extreme and presume that the two solutions
are essentially  just different pictures  of the same
physical situation, then it follows that Case $B$
should have a perfectly valid description in static
coordinates (namely, the $t=0$ spatial slice).
 Note that the compactness of the $t=0$ spatial section,
as implied by earlier analysis, is not of  issue in this analogy;
we can make the volume of this section  arbitrarily  
large by increasing the  ``width'' parameter, ${\cal A}$.
\par
On the other hand, it is not quite clear, at least in this exotic case, how
literally one should take the definition of forward time
as dictated by the implied  holographic duality.
In fact, the  relevant  ``$C$-theorem''  \cite{str2,bdm,lmmx} 
assumes that any additional  matter  satisfies $\rho>0$
and $\rho>|p|$ (i.e., the weak energy condition).  
By hypothesis, Case $B$ fails the
first of these conditions, and so we could have anticipated
the unorthodox outcome {\it a priori}.
Perhaps, there should be an analogous $C$-theorem for
negative-energy scenarios whereby the flow directions are reversed.
To put it another way, what exactly constitutes the direction of
time flow, in any model exhibiting time-reversal symmetry,
is not particularly clear.

\section{Conclusion}

In summary, we have been studying  a toy cosmological model
that describes a ``perturbed'' de Sitter spacetime.  More
specifically, we have considered  some of the implications
of incorporating matter  into  an otherwise purely
de Sitter spacetime of arbitrary dimensionality. 
This ``supplementary'' matter
was assumed to  have a perfect-fluid description,  have a fixed
equation of state, and
 satisfy the  causal energy condition. On a more generic note, we
allowed the matter  to have  either a  positive (Case $A$)
or negative (Case $B$) energy density.
Although  the latter case describes exotic matter in
violation of  the null energy condition,  we  noted
the following  alternative interpretation:
 the ``subtraction'' of positive-energy matter from
an otherwise purely de Sitter spacetime.
\par
After some preliminary background and analysis,
we were able to derive an analytical  expression for
 the  metric   in  both of the pertinent cases. (Here,
we employed a methodology that was first used
by McInnes in a related work \cite{mci}.) 
To obtain these precise  formulations, it was necessary 
to establish some boundary conditions  on
our FRW cosmological framework. In this regard,
we proposed  a universe that is asymptotically
de Sitter in the distant past and, barring singularities,
AsdS in the far future. We also stipulated
that the constant-time spatial slices  are compact and flat.
Although this choice of   slicing was motivated, 
in large part, by convenience, we  also provided
arguments on the basis of  observational and holographic considerations.
Ultimately, we were able to show that both solutions exhibited
the correct asymptotic behavior.
\par
With the   solutions of interest explicitly realized, we went on to consider 
the cosmological implications of these outcomes. First of all,
the positive-energy
case ($A$) was shown to have an unavoidable  singularity
at $t=0$. We identified this singularity as being
a  ``big crunch'' (or a ``big bang'' in the time-reversed
scenario).  We then argued that such singular behavior
can be viewed  as a direct manifestation  of violating  certain holographic
 bounds.  It is significant that these bounds place an upper limit  
on the accessible entropy and total mass of any spacetime
with a positive cosmological constant \cite{bou,bdm}.
Conversely, the negative-energy case ($B$) turned out
to be regular throughout the spacetime manifold.  This was
not unexpected, inasmuch as  matter of this nature does not jeopardize
the holographic bounds.  A perhaps less obvious outcome
is that such a spacetime can  provide  a  concrete realization
of an arbitrarily ``tall'' and ``wide'' AsdS-universe \cite{lmmx}.
This was shown to be possible, in spite of  Case $B$
clearly violating the null energy condition and (therefore)
nullifying  a  critical   
 antecedent of the  tall-universe construction \cite{gao}.
\par
The  final phase of our analysis  focused  
on  holographic renormalization group flows.
Such RG flows are expected to be induced
on  AsdS-spacetime  boundaries as 
 a consequence
of the (conjectured) dS/CFT duality \cite{str}.
For both cases of interest,  we
employed a standard prescription \cite{str2} to calculate
 a generalized $C$-function. Moreover, we  tested  the resultant functions
to determine if they describe  RG flows that conform
to   the  relevant  $C$-theorem  \cite{bdm,lmmx}.
It was ultimately demonstrated that  the positive-energy case ($A$)
could indeed satisfy the pertinent theorem. On the other hand,
  time evolution in Case $B$ induced
a RG flow  in the ``wrong'' direction. (On the plus side
of the ledger, the Case-$B$ flow was shown to be  monotonic 
up to an  anticipated ``bounce'' at $t=0$.)
\par
Although  Case $B$ essentially failed its holographic ``litmus
test'', we did express a couple of salvational viewpoints.
Firstly, we  conjectured  that the Case-$B$ solution -
specifically, the $t=0$ spatial slice -
can effectively  be  viewed as an eternally  static universe. 
 The premise underlying this proposal is
 as follows. (i) The $t=0$ surface   describes an infinitely 
ultraviolet
fixed point along the RG trajectory.  (ii)
There  could well be a strong link between this spacetime
and a (static) Schwarzschild-de Sitter black hole,
considering that  both solutions represent
a negative-energy excitation of  pure de Sitter space 
\cite{bdm}. Secondly, we suggested that the usual
convention for defining  time evolution in an AsdS universe
(towards the UV/IR in an expanding/contracting phase) 
 may not be applicable to 
this model  or, in fact, any model that  violates the  weak
energy condition. Certainly, the ``arrow of time'' seems 
a rather vague notion  when the universe  has
an inherent time-reversal symmetry.
\par
Regardless of one's personal viewpoint,
a better understanding of this ``holographic  failure''
would seem  to be in order. One possible 
avenue would be an investigation into
 how this perplexing  behavior 
is encoded in the holographically  dual theory.
In this regard, it is worth noting that the
so-called ``Casimir entropy''
of a  dual CFT \cite{ver} can also be interpreted
as a generalized $C$-function \cite{kpsz,hal}.
Since the  Casimir entropy is a direct property of 
 a  boundary theory, unlike  our prior prescription for $C$ (\ref{88}),
 it is perhaps the
more ``fundamental'' description of a given  $C$-function. 
That is to say, the Casimir entropy may serve 
as a  better indicator of  how  the time arrow should
be orientated.  Unfortunately, it is not so  clear how one
goes about evaluating  the Casimir entropy
when the bulk theory is intrinsically non-static.  
One possibility may be to adapt a ``mutual 
entropy'' approach that has recently been advocated 
for the (total) CFT entropy
\cite{klxxx}.
\par
In conclusion,  the results of this paper establish,
if nothing else, that  it is no trivial task to
construct an AsdS universe that is both regular throughout
the manifold
and holographically well understood. To achieve this,
it would  appear that significant anisotropy
is required in the matter fields;  which  would, of course, be
expected of any halfway  realistic   model.
It would be interesting to extend the current treatment
(i.e.,  the treatment of McInnes \cite{mci})
in this sort of direction,
although numerical analysis would almost  certainly be required. 
Finally, let us note that a ``satisfactory'' model (with regard to
the above criteria)
has indeed been formalized \cite{lmmx}. However,  the ``matter''
was  introduced via a  dilatonic scalar along with a rather
extravagant form of  potential. 
It may still be of interest to construct a  suitable AsdS
model with matter of a more conventional nature.

\section{Acknowledgments}
\par
The author  would like to thank  V.P.  Frolov  for helpful
conversations.



\end{document}